\documentclass{ws-p8-50x6-00}

\newcommand{\BE}{\begin{eqnarray}}
\newcommand{\EE}{\end{eqnarray}}
\newcommand{\BEn}{\begin{eqnarray*}}
\newcommand{\EEn}{\end{eqnarray*}}
\newcommand{\barr}{\begin{array}} 
\newcommand{\earr}{\end{array}}
\newcommand{\bit}{\begin{itemize}}      
\newcommand{\eit}{\end{itemize}}
\newcommand{\bfl}{\begin{flusleft}}
\newcommand{\efl}{\end{flusleft}}
\newcommand{\bfr}{\begin{flushright}}
\newcommand{\efr}{\end{flushright}}

\newcommand{\bc}{\begin{center}}
\newcommand{\ec}{\end{center}}

\newcommand{\ben}{\begin{enumerate}}    
\newcommand{\een}{\end{enumerate}}

\begin{document}

\title{Noise-driven Synchronization in \\ Coupled Map Lattices}

\author{L. Baroni $^{(1),(2)}$, R. Livi$^{(1),(2)}$, 
         and A. Torcini$^{(2),(3)}$\\[-2mm]$ $}

\address{$^{1}$
        Dipartimento di Fisica, Universit\'a di Firenze, 
        L.go E. Fermi, 5 - I-50125 Firenze, Italy\\
        $^{2}$
        Istituto Nazionale di Fisica della Materia, UdR Firenze,
        L.go E. Fermi, 3 - I-50125 Firenze, Italy\\
        $^{3}$
        Dipartimento di Energetica, Universit\'a di Firenze
        via S. Marta, 3 - I-50139 Firenze, Italy}

\maketitle

\abstracts{Synchronization is shown to occur in spatially extended
systems under the effect of additive spatio-temporal noise.
In analogy to low dimensional systems, synchronized
states are observable only if the maximum Lyapunov exponent 
$\Lambda$ is negative. However, a sufficiently high noise 
level can lead, in map with finite domain of definition, to
nonlinear propagation of information, even in 
non chaotic systems. In this latter case 
the transition to synchronization is ruled 
by a new ingredient : the propagation velocity 
of information $V_F$. As a general statement, we can
affirm that if $V_F$ is finite 
the time needed to achieve a synchronized trajectory grows 
exponentially with the system size $L$, while it increases
logarithmically with $L$ when, for sufficiently large 
noise amplitude, $V_F = 0$ .
}

\section{Introduction}

Noise-driven phenomena like diffusion and growth processes, 
front propagation and interface dynamics are quite common 
and extensively studied problems
in equilibrium and non-equilibrium statistical mechanics~\cite{surface}.
Usually, they are described in terms of stochastic PDE's,
or probabilistic automata, where some random process is
introduced as a model of white thermal noise~\cite{noise,Kapral}. 

More recently, noise has been shown to be able also to control
chaotic fluctuations in low-dimensional dynamical systems,
yielding synchronization of replicae driven by the same
stochastic process. This problem has been widely 
investigated~\cite{lowd,HF,Lai,pikov} considering two 
chaotic maps, e.g. the logistic
one or the Lorenz system, coupled through the same additive
noise.
For sufficiently large noise amplitude,
the invariant measure of both maps can be modified in such a
way to favour the stabilization on the same stochastic orbit.  
Far from misterious, this effect can be explained and quantitatively
predicted by observing that, for some noise amplitude, the
Lyapunov exponent associated to the 
dynamics becomes negative, as brillantly argued by Pilovsky~\cite{pikov}. 
Clearly, this can happen only if the map dynamics has expanding and
contracting regions, so that noise may amplify the role of the
latter against the former by properly modifying the invariant measure
of each map~\cite{HF}. In fact, any synchronization effect is obviously 
absent in uniformly hyperbolic maps, e.g.
the Bernoulli shift.

In this paper we aim to extend this analysis to 
spatially extended dynamical systems under the influence of 
spatio-temporal noise. In particular, we have focused our attention
on some specific coupled map lattice (CML) models~\cite{cml}, that,
in the absence of noise, exhibit dynamical phases characterized
by robust statistical properties. 

The main issues contained in this manuscript are hereafter
summarized:

\begin{itemize}

\item{} A necessary condition for obtaining synchronization
in CML models is that the maximum Lyapunov exponent $\Lambda$
becomes negative: this extends the validity of Pikovsky's criterion
to spatially extended systems~\cite{pikov}.

\item{} Additive spatio-temporal noise introduces
effective discontinuities in coupled interval maps, 
that, for $\Lambda < 0$, yield "stable-chaotic" 
phases~\cite{CoupledMaps}.
In this case synchronization is ruled by a further
indicator, the propagation velocity of information
$V_F$~\cite{kan2}.

\end{itemize}

\section{Models and Tools}

The general model of CML driven by the same realization of
spatio-temporal noise is defined by the 
the following two-step evolution rule:
\begin{eqnarray}
& {\tilde x}_i^t &= (1 - \varepsilon) x_i^t + \frac{\varepsilon}{2} 
(x_{i-1}^t + x_{i+1}^t)  \cr
& x_i^{t+1} &=  f({\tilde x}_i^t)\,+\, \sigma\cdot \eta_i^t  
\label{map}
\end{eqnarray}
The real state variable $x_i^t$ depends on the discrete space and time
indices $i = 1,2,\cdots , L$ and $t = 1,2,\cdots,T $, respectively.
The diffusive coupling $\varepsilon$ varies in the interval [0,1], 
$\sigma$ is the amplitude of the normalized space-time dependent 
random variable $\eta_i^t$, uniformly distributed in the interval [-1,1],
and $f(x)$ is a mapping from ${\cal S} \in$ R onto the interval ${\cal I}
\in $R. 

\begin{figure}[t]
\begin{center}
%\figurebox{20pc}{15pc}{} % to have a box alone
\epsfxsize=15pc % will enlarge or reduce the postscript figures based on the xsize
\epsfbox{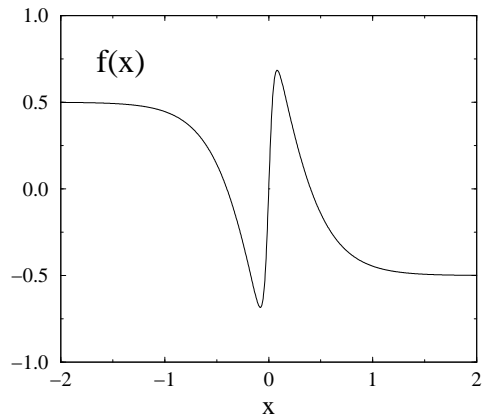} % postscript image file name
\end{center}
\caption{ The map $f(x)=\tanh(Ax)-B\tanh(Cx)$ for $A=20$, $B=1.5$
and $C=2$. \label{f01}}
\end{figure}

Following Ref.~\cite{Lai}, one can consider symmetric maps of 
the form shown in Fig.~\ref{f01}, 
\be
f(x)=\tanh(Ax)-B\tanh(Cx) 
\label{map1}
\end{equation}
with $A=20$, $B=1.5$ and $C=2$ and
where ${\cal S} \equiv R$ and ${\cal I} = [-a,a]$ ($a \simeq 0.684$). 
The presence of the noise term of amplitude $\sigma$ will obviously
enlarge the support of the invariant measure to the interval
$[-a-\sigma,a+\sigma]$. This will be not a problem for map (\ref{map1})
since ${\cal S}$ is thw whole real axis.
So far, the corresponding CML model is certainly not 
one of the most investigated problems in the literature. 
In fact, CML are usually defined using maps
$f$ of the interval [0,1] onto itself, like the logistic map,
that are known to exhibit specific dynamical features,
yielding many interesting statistical and dynamical effects in
the presence of diffusive coupling (e.g., see Ref.~\cite{Kapral,Kaneko,Chate}).
In this case ${\cal I} \equiv {\cal S} = [0,1]$ and the dynamic
rule (\ref{map}) can maintain the state variable $x_i^t$ inside 
${\cal S}$ only by adopting some further recipe: for instance
$x_i^{t+1} \rightarrow x_i^{t+1} + 1$  
( $x_i^{t+1} \rightarrow x_i^{t+1} - 1$) 
if  $x_i^{t+1} < 0 $ ($x_i^{t+1} > 1$).
This rule modifies the probability measure
of the state variable $x_i^t$ w.r.t. the noise-free case. 
Herzel and Freund~\cite{HF} showed that, in low-dimensional
systems, a similar recipee breaks the symmetry of the noise
process, giving rise to an effective state-dependent noise, 
with nonvanishing mean. 
Later, Lai and Zhou~\cite{Lai} 
have shown that, even maintaining symmetric noise, synchronization
can be achieved, provided sufficiently large noise amplitudes 
amplify the role of contracting regions in phase space.

This notwithstanding, for what concerns synchronization in
high dimensional dynamical systems like CML, it is more 
interesting to point out that the above mentioned recipee 
introduces effective discontinuities. In fact, two state variables,
whose values are close to one another and to one of the boundaries,
can be driven by the addition of spatio-temporal noise at a 
distance of order 1. This amounts to a strong non-linear effect,
that is highly reminiscent of the mechanism yielding stable
chaos in CML (see Ref.~\cite{CoupledMaps}). 
Here it is induced by noise,
while in the just quoted CML model the discontinuity was
built in the map $f$. As we are going to show in Sec. 3,
the only distinctive feature of these
scenarios is that the final attractor is a stable periodic 
orbit in the deterministic CML and a stochastic one, in the
noisy model.

In the following section we discuss synchronization in both
of the above mentioned models of noise-driven CML. For this
purpose, here it is worth defining the main quantitative
indicators that we have used for working out our analysis.
A first sensible quantity is the maximum Lyapunov exponent, 
$\Lambda$, measured according to
the method outlined in Ref.~\cite{benettin}. We want to remark
that this quantity is defined for deterministic dynamics (i.e.
the noise-free case, $\sigma = 0$ ) and it measures the average 
exponential expansion rate of nearby trajectories within a linear 
stability analysis.
For $\sigma \not= 0$ we are faced with stochastic trajectories and it is not 
{\sl a priori} obvious that $\Lambda$ is still a well-defined quantity. 
On the other hand, the linearized evolution equations in the tangent space
of Eq.~(\ref{map}) do not depend explicitely on
$\eta_i^t$. Accordingly, they are formally equivalent for both the noisy
and the noise-free cases. However, the presence of noise modifies the
dynamical evolution of the system and therefore also the tangent
space dynamics. Careful numerical simulations show 
that $\Lambda$ remains a well defined asymptotic quantity, 
measuring the exponential expansion rate
of infinitesimal perturbations of the stochastic evolution.
In particular, its value is found to depend on $\sigma$, but not
on the realization of noise.
 
Another relevant indicator for the study of spatially extended dynamical 
systems is the average propagation velocity of finite amplitude 
perturbations~\cite{kan2}
\be
V_F 
= \lim_{t\to\infty} \lim_{L\to\infty} 
\frac{\langle F(t)\rangle}{t}
\label{eq:velf}
\ee
where 
\be
F(t) =
\max \bigg\{i : 1 \leq i \leq L \, ;\, |x_i^t - y_i^t| > 0 \bigg\} \quad.
\label{front}
\ee
The trajectories $x_i^t$ and $y_i^t$ represent the synchronous evolution
according to Eq.~(\ref{map}), with the same realization of noise $\eta_i^t$,
of two initial conditions that differ at time $t=0$ for finite
amounts $ \delta_i \sim {\cal O}(1)$ only inside a space region of 
size $S$:
$y_i^0 = [x_i^0 + \delta_i , {\rm mod} 1]$ for $ |L/2-i| \leq S$, otherwise
$y_i^0 = x_i^0 $~.

In the noisy model the average in (\ref{eq:velf}) has to be performed 
over different initial conditions and noise realizations.
It is worth stressing that $V_F$ is associated
to the nonlinear mechanisms of information propagation and 
it can have a finite value even for non chaotic 
evolution~\cite{CoupledMaps}.
Accordingly, in order to have no information flow
in a spatially extended system not only $\Lambda$ 
but also $V_F$ should vanish.

\section{Synchronization of Replicae}

Two natural quantities can be introduced for identifying the synchronization
of replicae: the {\sl first passage time} $\tau_1(\Delta)$, i.e.
the time needed for two orbits, starting from different initial conditions 
$x_i^0$ and $y_i^0$, to approach each other at a distance 
$ d(t) \, =\, \frac{1}{N} \sum_{i=1}^N |x_i^t - y_i^t|$
smaller than a given threshold $\Delta $ (usually assumed much smaller
than unity); the time 
$\tau_2(\Delta)$ during which their distance
remains smaller than $\Delta$~\cite{nota1}.
Since, in general, these times should depend on the initial 
conditions and on the noise realizations, 
in what follows we shall use the 
same symbols for denoting the averaged quantities. 
In all the numerical calculations hereafter reported we have
used $\Delta \sim 10^{-8} - 10^{-10}$. We have also verified that 
results are not affected by the choice of $\Delta$,
provided it is small enough, tipically $\Delta < 10^{-7}$.

\begin{figure}[t]
\begin{center}
%\figurebox{20pc}{15pc}{} % to have a box alone
\epsfxsize=15pc % will enlarge or reduce the postscript figures based on the xsize
\epsfbox{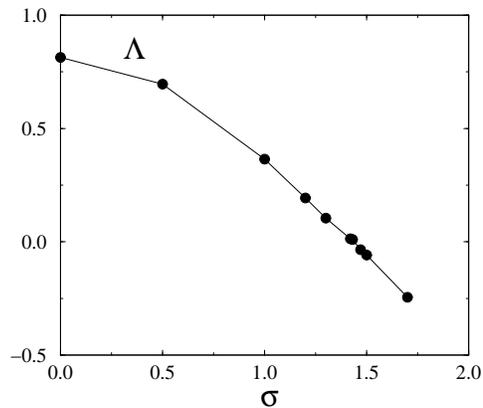} % postscript image file name
\end{center}
\caption{Lyapunov exponent $\Lambda$ as a function of the amplitude
of the noise $\sigma$ for a chain of $1000$ 
coupled maps of the type reported in Fig.~\ref{f01} for $\varepsilon = 0.1$.
The $\Lambda$-values have been obtained by integrating for 
$130,000$ time steps.  \label{f02}}
\end{figure}

\noindent
As a first example let us consider the CML with the map
$f$ shown in Fig. ~\ref{f01}. In this case, we have
observed synchronization for all the $\varepsilon$-values
that we tested.

For instance, with a diffusive coupling value 
$\varepsilon = 0.1$ we have found that synchronization is 
obtained for $\sigma > \sigma_s = 1.44$. 
In Fig. ~\ref{f02} we show that $\Lambda$ 
is a function of $\sigma$ and it becomes negative at $\sigma_s$.
This is the first evidence of the validity of
Pikovsky's criterion also for spatially extended systems.

In principle, one has to consider that $\tau_1(\Delta)$ 
is also a function of the system size $L$.
For $\sigma < \sigma_s$, i.e. $\Lambda > 0$, we find that 
$\tau_1(\Delta)$ diverges exponentially with the 
system size $L$ : $\tau_1(\Delta) \sim e^{L/\xi}$.
The scale factor $\xi(\varepsilon,\sigma)$ plays the role
of an effective correlation length and
is found to be finite for any choice of the parameters~\cite{nota2}.
This exponential divergence can be explained by a simple argument.
Despite the spatial coupling, the combined effect of chaos 
and noise makes an effective number
of degrees of freedom, $L/\xi$, in the replicae evolve independently.
Accordingly,
the probability that the replicae get closer than a distance
$\Delta$ should be proportional to $\Delta^{L/\xi}$ and 
$\tau_1(\Delta)$ can be reasonably assumed to be 
inversely proportional to this probability.

\begin{figure}[t]
\begin{center}
%\figurebox{20pc}{15pc}{} % to have a box alone
\epsfxsize=15pc % will enlarge or reduce the postscript figures based on the xsize
\epsfbox{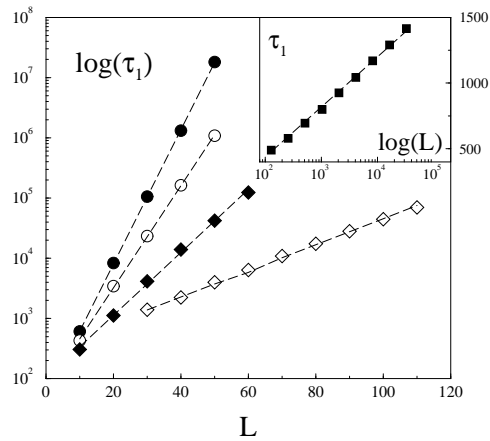} % postscript image file name
\end{center}
\caption{ Exponential scaling of $\tau_1$ versus $L$ for
coupled logistic maps with $\varepsilon = 2/3$ at
different values of the noise amplitude: $\sigma$ = 0.3 (empty circles),
0.32 (filled circles), 0.35 (empty diamonds), 0.38 (filled diamonds).
The inset shows the logarithmic scaling of $\tau_1$ versus $L$ 
for $\sigma = 0.45$~. The values of $\tau_1$ have been computed
with $\Delta = 10^{-10}$ and averaged over
$10^3$ initial conditions. \label{f2}}
\end{figure}

At variance, a logarithmic dependence of $\tau_1(\Delta)$ on
$L$ is found for $\sigma > \sigma_s$.
Numerical simulations show that the number of 
regions made of a few synchronized sites increases as time flows.
Moreover, once a synchronized region is formed, it
grows linearly in time, until all regions merge and 
synchronization sets in over the whole lattice. Upon these observations,
one can introduce a simple model accounting for 
the logarithmic dependence of $\tau_1$ on $L$.
An effective rate equation for the number of synchronized sites,
$n(t)$, can be constructed by assigning a
probability $p$ for the formation of new synchronized sites and
a rate $\gamma$ for the linear increase of synchronized regions:
\begin{equation}
\label{eqrate}
\frac{\partial n}{\partial t} =
\gamma + p(L-n) \quad ,
\end{equation}
with $0 \le n(t) \le L$. Solving this equation with the initial
condition $n(0)=0$, one obtains an estimate of 
$\tau_1$ by imposing the condition $n(\tau_1) = L$:
\begin{equation}
\label{tau1}
\tau_1 = \frac{1}{p} \ln \left[ \frac{pL}{\gamma} +1
\right] \quad .
\end{equation}
Note that a logarithmic dependence with $L$ (see, e.g., the inset
of Fig. ~\ref{f2}) is consistent with the condition $\frac{pL}{\gamma} >> 1$,
that can be always achieved for sufficiently large values of $L$.

For what $V_F$ is concerned, in this case it is not expected to 
provide any additional information w.r.t. $\Lambda$. 
Actually, $V_F$ is found to be positive for $\sigma < \sigma_s$, 
while it vanishes for $\sigma > \sigma_s$.

Different scenarios are obtained by considering the second
kind of noisy CML described in the previous section.
Here we consider the logistic map at the Ulam point
\begin{equation}
f(x) = 4 x (1-x)
\label{logistic}
\end{equation}
Let us first study the case $\varepsilon =1/3$:
for large enough values of $L$, 
$\Lambda$ and $V_F$ are positive for any $\sigma$.
However, their values are found to vary with $\sigma$,
as well $\tau_1$ and $\tau_2$ that 
remain finite up to $\sigma \sim {\cal O}(1)$.
Again, $\tau_1$ is found to diverge exponentially with $L$.

\begin{figure}[t]
\begin{center}
%\figurebox{20pc}{15pc}{} % to have a box alone
\epsfxsize=17pc % will enlarge or reduce the postscript figures based on the xsize
\epsfbox{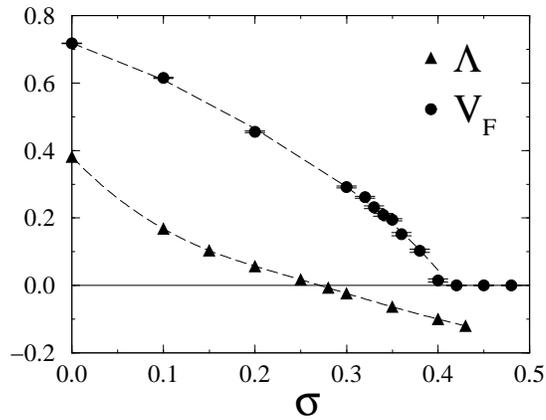} % postscript image file name
\end{center}
\caption{Behaviour of the propagation velocity $V_F$ (circles) and
of the maximum Lyapunov exponent $\Lambda$ (triangles) versus  
noise amplitude $\sigma$ for coupled logistic maps with
$\varepsilon = 2/3$. Both quantities have been computed for $L=1024$,
averaging over $10^3$ initial conditions each one followed for 
${\cal O}(10^5)$ time steps.
The dashed lines are a guide for the eyes.  \label{f1}}
\end{figure}

A new interesting scenario occurs for $\varepsilon = 2/3$, where, 
for $L \geq {\cal O} (10)$,  
$\Lambda$ is negative for $\sigma > 0.27$ (see Fig.~\ref{f1}). 
Therefore, one expects that synchronization occurs and this is
indeed the case although, 
for $ 0.27 < \sigma < 0.4$~,
$\tau_1(\Delta)$ is still found to increase exponentially with $L$
(see Fig. ~\ref{f2}). 
On the other hand, for $\sigma > 0.4$  $\tau_1(\Delta)$
grows logarithmically with $L$ (see the inset of Fig.~\ref{f2}).

\begin{figure}[t]
\begin{center}
%\figurebox{20pc}{15pc}{} % to have a box alone
\epsfxsize=15pc % will enlarge or reduce the postscript figures based on the xsize
\epsfbox{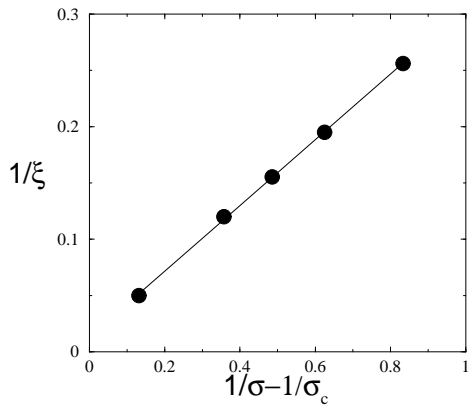} % postscript image file name
\end{center}
\caption{ Inverse of the scale factor $1/\xi$ as a function of
$\sigma^{-1} - \sigma_c^{-1}$ ($\sigma_c = 0.4$) for CML logistic maps with
$\varepsilon = 2/3$ and $0.30 \le \sigma \le 0.38$ . \label{f3}}
\end{figure}

This transition between two completely different
dynamical regimes is ruled by $V_F$, that drops to zero at 
$\sigma_c \approx 0.4$~(see Fig. ~\ref{f1})~. In fact, for $\sigma < \sigma_c$,
a positive $V_F$ implies that any perturbation of finite
amplitude propagates through the lattice. Despite the linear
mechanism is no more active ($\Lambda < 0$), we find that the
non-linear mechanism of information propagation 
maintains the exponential dependence of $\tau_1$ on $L$,
thus still implying that an effective number of degrees of freedom 
evolves independently. In particular, we find numerical evidence
of the relation $\xi^{-1} \propto (\sigma^{-1} - \sigma_c^{-1})$ 
for $\sigma < \sigma_c$ (see Fig.~\ref{f3});  this indicates that the
scale  factor $\xi$ diverges at the transition
point.

Eventually, only when $d(t)$
becomes sufficiently small the linear mechanism acts as a stabilizing
factor on the dynamics of the replicae and synchronization is achieved.

For $\sigma > 0.4$, where also the nonlinear mechanism is absent,
numerical simulations exhibit the same scenario described in
the ``symmetric map'' CML  for $\sigma > \sigma_s$.

One can counclude that in this last case the
conditions $\Lambda < 0$ and $V_F = 0$ 
have to be fulfilled independently for guaranteeing 
noise-driven synchronization of two replicae within a ``reasonable'' 
time span in a CML of large but finite size $L$.

\begin{figure}[t]
\begin{center}
%\figurebox{20pc}{15pc}{} % to have a box alone
\epsfxsize=15pc % will enlarge or reduce the postscript figures based on the xsize
\epsfbox{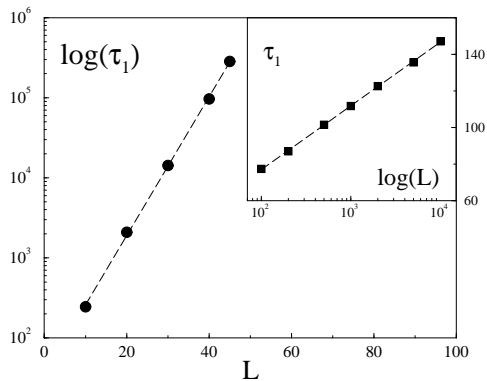} % postscript image file name
\end{center}
\caption{ Exponential scaling of $\tau_1$ versus $L$ for
coupled stable maps with $\varepsilon = 0.31$ and 
$\sigma$ = 0.15 (filled circles).
The inset shows the logarithmic scaling of $\tau_1$ versus $L$ 
for $\sigma = 0.25$ (filled squares). The values of $\tau_1$ have been computed
with $\Delta = 10^{-8}$ and averaged over
$10^3$ initial conditions. \label{f4}}
\end{figure}

We want to point out that this scenario is not peculiar of the 
logistic map for $\varepsilon = 2/3$. 
Actually, we have verified that it holds in a finite interval of values
around $\varepsilon = 2/3$. 
Moreover, a very similar situation is obtained when considering dynamics 
(\ref{map}) for a model of period-3 stable maps~\cite{CoupledMaps}
\begin{equation} 
   f(x)  = \cases{ bx  & $0 < x < 1/b$ \cr
                   a + c(x-1/b) & $ 1/b < x < 1$}
   \label{noi} 
\end{equation} 
where $b = 2.7$, $a=0.07 $ and $c=0.1$. This case is particularly
interesting for a twofold reason. 
It has been shown that the noise-free case has a negative $\Lambda$ 
for any value of $\varepsilon$
and, moreover, it exhibits a peculiar 
transition at $\varepsilon_c \approx 0.6$ from a frozen disordered 
phase with $V_F = 0$ to a chaotic phase with $V_F > 0$~\cite{cecconi}. 
At variance with the case of coupled logistic maps, when noise is added
$\Lambda$ remains negative for any value of $\sigma$. 
The phase transition disappears, because, independently of $\varepsilon $,
$V_F$ is found to be positive for very small noise amplitudes.
On the other hand, by increasing $\sigma$ up to a
critical value $\sigma_c(\varepsilon)$~, $V_F$ is found to drop again 
to zero not only 
below, but also above $\varepsilon_c$: for instance, one has $\sigma_c
\approx 0.16$ for $\varepsilon = 0.58$ and $\sigma_c\approx 0.18$ for
$\varepsilon = 0.62$.
In both of these cases we have recovered the 
same kind of mechanisms characterizing the synchronization transition 
discussed for the case of coupled logistic maps with $\varepsilon = 2/3$
(see Fig. ~\ref{f4}).
In this sense, it seems reasonable to conjecture that this phenomenon 
is present
in a wide class of spatially extended dynamical systems and that it can 
be characterized making use of the same kind of analysis discussed in
this paper for CML models. 

As a final remark, we would like to stress that all this scenario 
is reminiscent of the phenomenology
associated to stable chaos in CML, that was identified for
model~(\ref{noi})~\cite{CoupledMaps}. 
Also in that case unpredictability was characterized by 
transients increasing exponentially with $L$ when $V_F > 0$.
The analogy is strenghtened by the fact that additive noise
induces ``jumps'' in the deterministic dynamics of the system,
similar to the discontinuities present in model (\ref{noi}).
As a matter of fact, the presence of strong
variations in the slope of the local map is fundamental
in order to observe non-linear propagation mechanisms~\cite{grass}.

\section*{Acknowledgments}

We want to thank F. Bagnoli and A. Politi for helpful discussions.
Part of this work was performed at the Institute of Scientific Interchange
in Torino, during the workshop on `` Complexity and Chaos '', June 1998. 
We acknowledge CINECA in Bologna and INFM for providing 
us access to the parallel computer CRAY T3E under the  
grant `` Iniziativa Calcolo Parallelo''.


\begin{thebibliography}{99}
\bibitem{nota1}  We have checked that different norms, e.g.
the maximum or the euclidean ones, do not modify the resulting scenario.
Note that both quantities are well defined also 
in the noise-free case, since there is a finite 
probability that two different initial conditions may stay closer than 
a finite distance $\Delta$.
\bibitem{nota2} Although intricate, the dependence 
of $\xi$ on the parameters does not add any relevant feature
to the outlined scenario. For the sake of space it will be discussed
in a forthcoming paper, containing more details and more examples than
those reported in this letter.
\end{thebibliography}
\end{document}